\documentclass[aps,notitlepage,nofootinbib]{revtex4-1}
\usepackage{epsfig}
\usepackage{color}
\usepackage{url}
\usepackage{amsmath}
\usepackage{multirow}
\def\bea#1\eea{\begin{align}#1\end{align}}

\newcommand{\bef}{\begin{figure}[htb]\centering}
\newcommand{\eef}{\end{figure}}

\begin{document}
\title{Photon-tagged and B-meson-tagged b-jet production at the LHC}

\date{\today}

\author{Jinrui Huang}
\email{jinruih@lanl.gov} 
\affiliation{Theoretical Division, 
                   Los Alamos National Laboratory, 
                   Los Alamos, NM 87545, USA}

\author{Zhong-Bo Kang}
\email{zkang@lanl.gov}
\affiliation{Theoretical Division,
                   Los Alamos National Laboratory,
                   Los Alamos, NM 87545, USA}
                   
\author{Ivan Vitev}
\email{ivitev@lanl.gov}                   
\affiliation{Theoretical Division, 
                   Los Alamos National Laboratory, 
                   Los Alamos, NM 87545, USA}       
                   
\author{Hongxi Xing}
\email{hxing@lanl.gov}
\affiliation{Theoretical Division, 
                   Los Alamos National Laboratory, 
                   Los Alamos, NM 87545, USA}                                      

\begin{abstract}
 Tagged jet measurements in high energy hadronic and  nuclear reactions  provide constraints on 
the energy and parton flavor origin of the parton shower that recoils against the tagging particle. 
Such additional insight can be especially beneficial in illuminating the mechanisms of heavy flavor 
production in proton-proton collisions at  the LHC and their modification in  the 
heavy ion  environment, which are not fully understood. With this motivation, we present theoretical
results for isolated-photon-tagged and B-meson-tagged b-jet production at $\sqrt{s_{NN}} = 5.1$ TeV 
for comparison to the upcoming lead-lead data.  We find that photon-tagged b-jets exhibit smaller momentum 
imbalance shift in nuclear matter, and correspondingly smaller energy loss, than photon-tagged light flavor jets. 
Our results show that B-meson tagging is most effective in ensuring that the dominant fraction of 
recoiling jets originate from prompt b-quarks. Interestingly, in this channel the large suppression of the cross section is 
not accompanied by a significant momentum imbalance shift.  
\end{abstract}

\maketitle

\section{Introduction}
High transverse momentum ($p_T$) production of energetic particles~\cite{Owens:1986mp}  and 
jets~\cite{Sterman:1977wj} in proton-proton (p+p), proton-nucleus (p+A) and 
nucleus-nucleus (A+A) collisions  plays an essential role in understanding the fundamental theory of strong interactions,
Quantum Chromodynamics (QCD), and in  probing the properties of cold nuclear matter and  the quark-gluon plasma (QGP) 
created in relativistic heavy-ion collisions at the Relativistic Heavy Ion Collider (RHIC) and the Large Hadron Collider 
(LHC)~\cite{Wang:1991xy,Gyulassy:2003mc}.  Recent theoretical and phenomenological advances, ranging from 
improvements in energy loss  models~\cite{Burke:2013yra} to a unified treatment of vacuum and  medium-induced 
parton showers~\cite{Kang:2014xsa}, have resulted in a good description of the light hadron cross sections attenuation
in heavy ion collisions relative to binary collision-scaled p+p result, a phenomenon known as jet quenching. 
Predictions for light flavor reconstructed jet suppression~\cite{He:2011pd} were also confirmed by LHC 
measurements~\cite{Aad:2012vca}. However, early expectations that the corresponding 
suppression of D-meson and B-meson production will be significantly smaller 
due to heavy quark mass and color charge effects~\cite{Dokshitzer:2001zm} were not supported by RHIC
data~\cite{Adare:2006nq,Abelev:2006db}. The observed discrepancy between theory and experiment has inspired 
the development of a variety of theoretical models~\cite{Wicks:2005gt,Adil:2006ra,vanHees:2007me,Qin:2009gw,Gossiaux:2008jv,Sharma:2009hn}. 
Consistent understanding of light and heavy flavor suppression still remains a challenge for theory in light of 
the  LHC  results~\cite{ALICE:2012ab}.

Heavy flavor jets at the LHC have emerged as a new tool to further test the theory of heavy 
quark production, parton shower  formation initiated by prompt b-quarks,  and the modification 
of these processes in nuclear matter due to energy loss effects. Detailed studies of b-jets may also 
become part of the RHIC experimental program in the future~\cite{Adare:2015kwa}.  In a previous 
work~\cite{Huang:2013vaa}, we studied single inclusive  b-jet production in heavy ion collisions. 
The predicted suppression of the b-jet cross section in lead-lead (Pb+Pb) reactions is consistent 
with subsequent experimental measurements by the CMS  Collaboration~\cite{Chatrchyan:2013exa} 
and supports a coherent energy loss picture, where heavy quark mass effects disappear at high 
$p_T$.  At the same time, the inclusive nature of this observable does not allow to extract
in a model-independent way the fraction of the b-jet energy that is dissipated in the medium.      
Furthermore, inclusive b-jets receive a large contribution from prompt light quarks and gluons, where 
heavy flavor emerges from parton splitting only in the late stages of the parton shower evolution. Thus, the
connection between b-jets suppression and b-quark energy loss can be quite indirect.

In this paper, our main goal is to identify differential observables that enhance the fraction of b-jets 
that  originate from prompt b-quarks and provide constraints on the energy of the parton shower.    
To this end, we extend our calculations of inclusive b-jets to isolated-photon-tagged and
B-meson-tagged b-jets production in heavy  ion collisions at center-of-mass energy per 
nucleon-nucleon pair $\sqrt{s_{NN}}=5.1$ TeV, to be 
compared with experimental data from the forthcoming Pb+Pb run at the LHC. We discuss the advantages of these final states 
for studying the physics of b-quark production and propagation in dense QCD matter and present 
quantitative  predictions for the tagged b-jet cross section suppression and the modification of the 
related momentum imbalance distribution in the QGP. 
We note, that earlier studies of photon-tagged~\cite{Stavreva:2009vi,Hartanto:2013aha,Kang:2011rt,Stavreva:2012aa}  and heavy 
meson-tagged~\cite{Vitev:2006bi,Zhu:2006er,Cao:2015cba}  heavy flavor  in p+p and A+A
reactions have also advocated the use of more exclusive final states to unravel the dynamics of
heavy flavor production in hadronic and nuclear reactions.

The rest of our paper is organized as follows. In Sec.~II we present the evaluation of the isolated-photon-tagged 
and B-meson-tagged b-jet cross sections in p+p collisions using Pythia 8 simulations. In Sec.~III we first describe 
the simulation of the medium-induced parton shower that recoils against the tagging particle and the related 
heavy quark energy loss effects. We then present our phenomenological results for the tagged b-jet production 
in Pb+Pb collisions at the LHC. We conclude our paper in Sec.~IV.

\section{Photon-tagged and B-meson-tagged b-jet production in p+p collisions}
\label{pp}
In this section, we present the evaluation of the differential cross sections for photon-tagged and B-meson-tagged b-jet production in p+p collisions using Pythia 8~\cite{Sjostrand:2007gs}. This generator utilizes leading-order perturbative QCD matrix elements for $2\to 2$ processes, combined with a leading-logarithmic $p_T$-ordered parton shower, and the Lund string model for hadronization. Specifically, we use the CTEQ6L1 parton distribution functions~\cite{Pumplin:2002vw} in the simulation and perform the jet clustering with the anti-$k_T$ algorithm~\cite{Cacciari:2008gp} and a jet radius parameter $R$. In the photon-tagged b-jet production simulation, we first select the photon according to the desired kinematics and isolation cut specified below, and then we identify the b-jet on the away side with $|\phi_\gamma - \phi_j| > 3/4 \pi$, where $\phi_\gamma$ ($\phi_j$) is the azimuthal angle of the photon (b-jet). The b-jets are defined to be jets that contain at least one b-quark (or $\rm \bar b$-quark) inside the jet cone: a b-quark (or $\rm \bar b$-quark) is assigned to a jet if the radial separation from the reconstructed jet axis satisfies $\Delta R < R$, where $\Delta R = \sqrt{(\Delta \phi)^2+(\Delta y)^2}$ with $\phi$ and $y$ being the azimuthal angle and  rapidity, respectively. For the B-meson-tagged b-jet production, which has a much larger cross section in comparison to
$\gamma$-tagged  b-jet production, we select only the leading B-meson in each event. 

\begin{figure}[!t]
\psfig{file=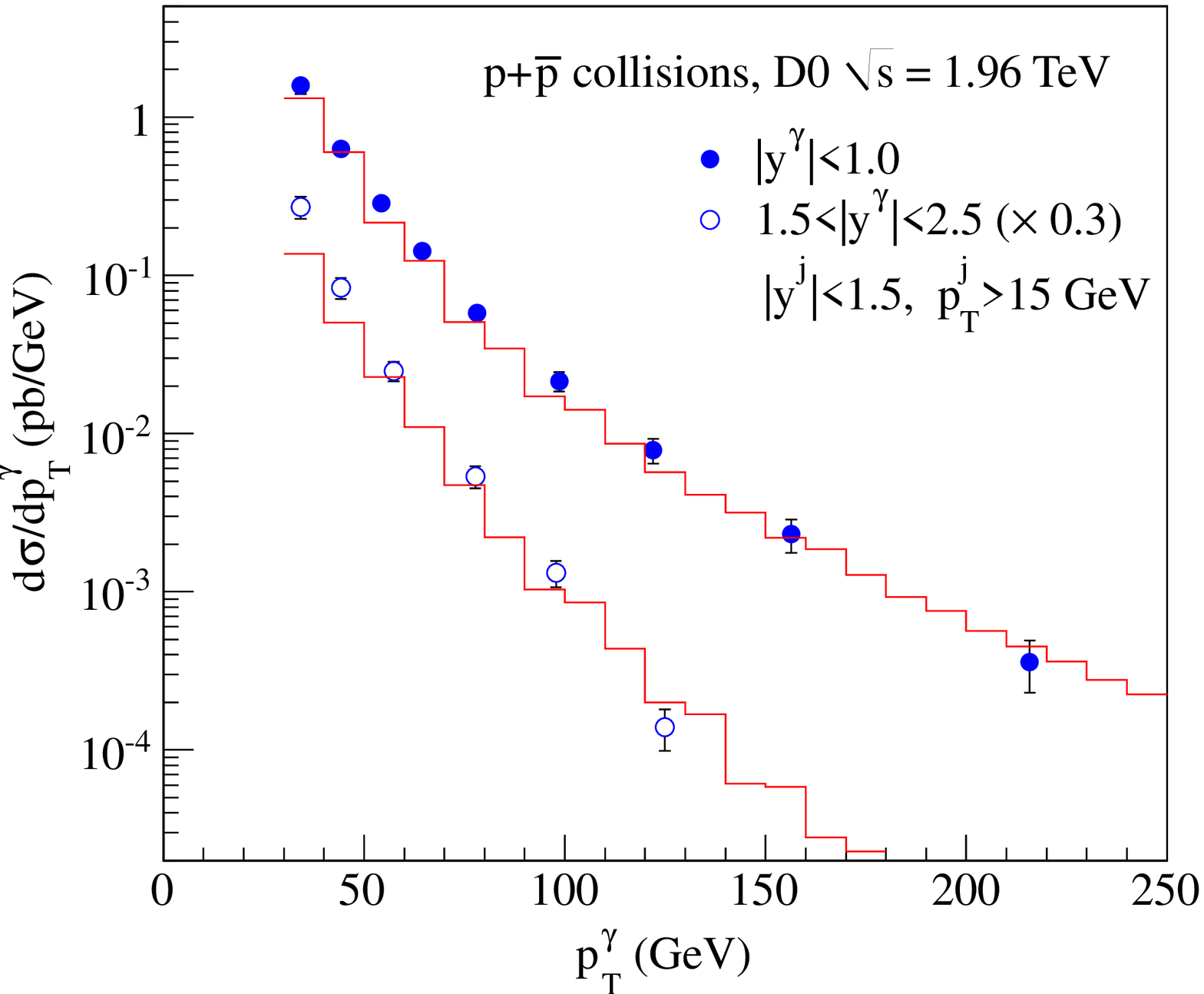, width=2.8in}
\hskip 0.2in
\psfig{file=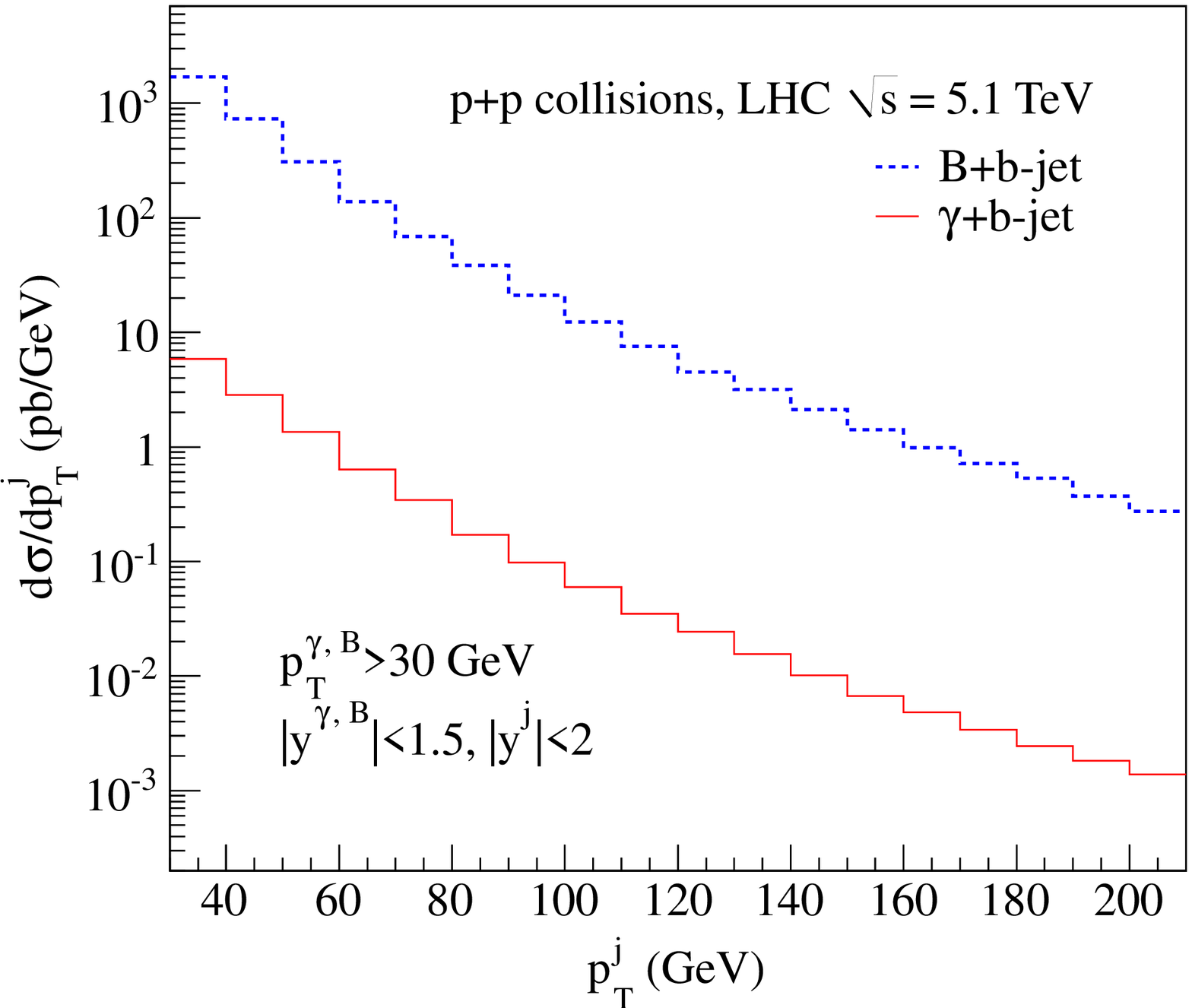, width=2.74in}
\caption{Left: The $\gamma$+b-jet differential cross sections in p+${\rm\bar p}$ collisions at the Tevatron $\sqrt{s}=1.96$ TeV is shown as a function of $p_T^\gamma$ in two photon rapidity regions, $|y^\gamma|<1.0$ (filled circles) and $1.5 < |y^\gamma| < 2.5$ (open circles, multiplied by 0.3 for presentation), as measured by the D0 Collaboration~\cite{Abazov:2012ea}. The b-jet kinematics cuts are $|y^j| < 1.5$ and $p_T^j > 15$ GeV. Red curves are from Pythia 8 simulations, using the anti-$k_T$ algorithm with $R=0.5$.
Right: The same differential cross section from Pythia 8 simulation in p+p collisions at LHC $\sqrt{s}=5.1$ TeV as a function of $p_T^j$. The tagging particles (either the leading B-meson or the isolated $\gamma$) have the following kinematic cuts $|y^{B,\gamma}|<1.5$ and $p_T^{\gamma, B} > 30$ GeV. The b-jets are constructed using anti-$k_T$ algorithm with $R=0.3$ and have rapidity $|y^j| < 2$. The blue (red) curve is for the $B$-meson-tagged (isolated-$\gamma$-tagged)  b-jet production. }
\label{fig:pp-cross}
\eef

Measurements of the production cross section of isolated photons associated with b-jets in proton-antiproton (p+$\rm \bar p$) collisions at $\sqrt{s} = 1.96$ TeV were performed by both D0~\cite{Abazov:2012ea} and CDF~\cite{Aaltonen:2013coa} Collaborations at the Tevatron. Our baseline simulation can, thus, be validated against the experimental data, as shown in Fig.~\ref{fig:pp-cross} (left). Here, the differential cross section $d\sigma/dp_T^\gamma$ is plotted as a function of photon transverse momentum $p_T^\gamma$. We implement the experimental cuts in our simulation: for the b-jets $|y^j| < 1.5$, $p_T^j > 15$ GeV, and the jet radius parameter $R=0.5$. For photons, two rapidity cuts are considered:  $|y^\gamma| < 1.0$ and $1.5 < |y^\gamma| < 2.5$. The photon isolation cut is also imposed in the simulation, which requires the hadronic transverse energy in a cone of radius $R_{\rm iso} = \sqrt{(y-y_\gamma)^2 + (\phi - \phi_\gamma)} = 0.4$ around the photon direction to be less than 7\% of the photon energy. As can be seen  in Fig.~\ref{fig:pp-cross} (left), our simulations give a good description of the D0 experimental data.

In the following section we will  present results on the nuclear modification of both isolated-photon-tagged and B-meson-tagged b-jet production in Pb+Pb collisions at $\sqrt{s} = 5.1$~TeV. We first discuss the baseline calculation, i.e.  the differential cross sections in p+p collisions at the same center-of-mass energy. In this simulation (and all the calculations below) we integrate over the b-jet rapidity $|y^j| < 2$ and the isolated-photon and B-meson rapidity $|y^{\gamma,B}| < 1.5$. With the transverse momentum $p_T^{\gamma, B} > 30$~GeV, the differential cross sections are plotted as a function of the b-jet transverse momentum $p_T^j$ in Fig.~\ref{fig:pp-cross} (right). As can be seen from the figure, the $\gamma$-tagged b-jet production cross section is comparable to the one measured at the Tevatron, while B-meson-tagged b-jet cross section can be more than two orders of magnitude larger. Both of those observables should be experimentally accessible at the LHC~\cite{ATLAS:2011ac}. We note that an uncertainty in the absolute cross sections, for example due to different Pythia tunes, will cancel in the nuclear modification which we study in the next section. To reduce the statistical uncertainty/fluctuations in the baseline p+p cross sections, we have simulated around $10^8$ events for both B-meson-tagged b-jets and $\gamma$-tagged b-jets.

\begin{figure}[!t]
\psfig{file=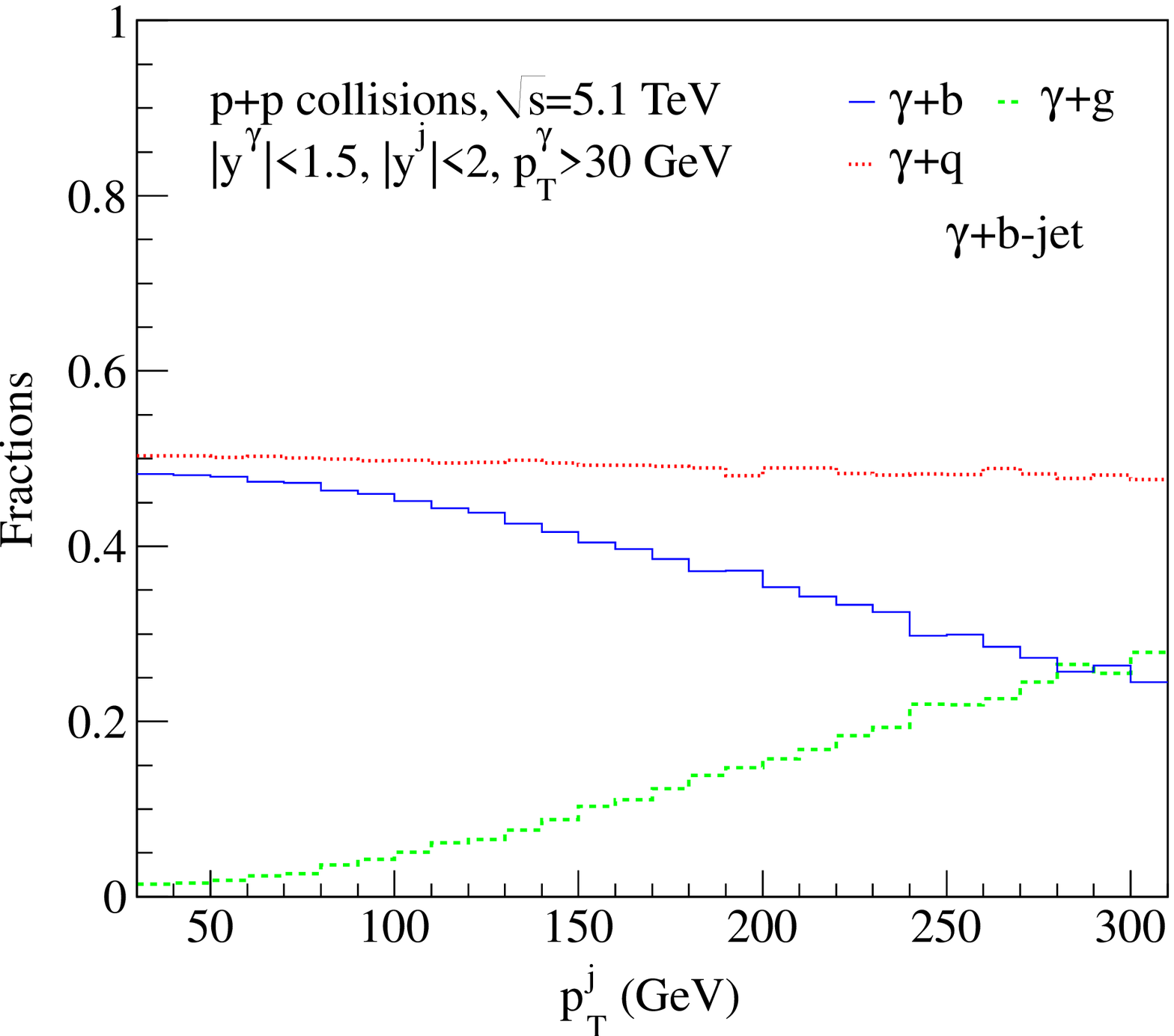, width=2.8in}
\hskip 0.2in
\psfig{file=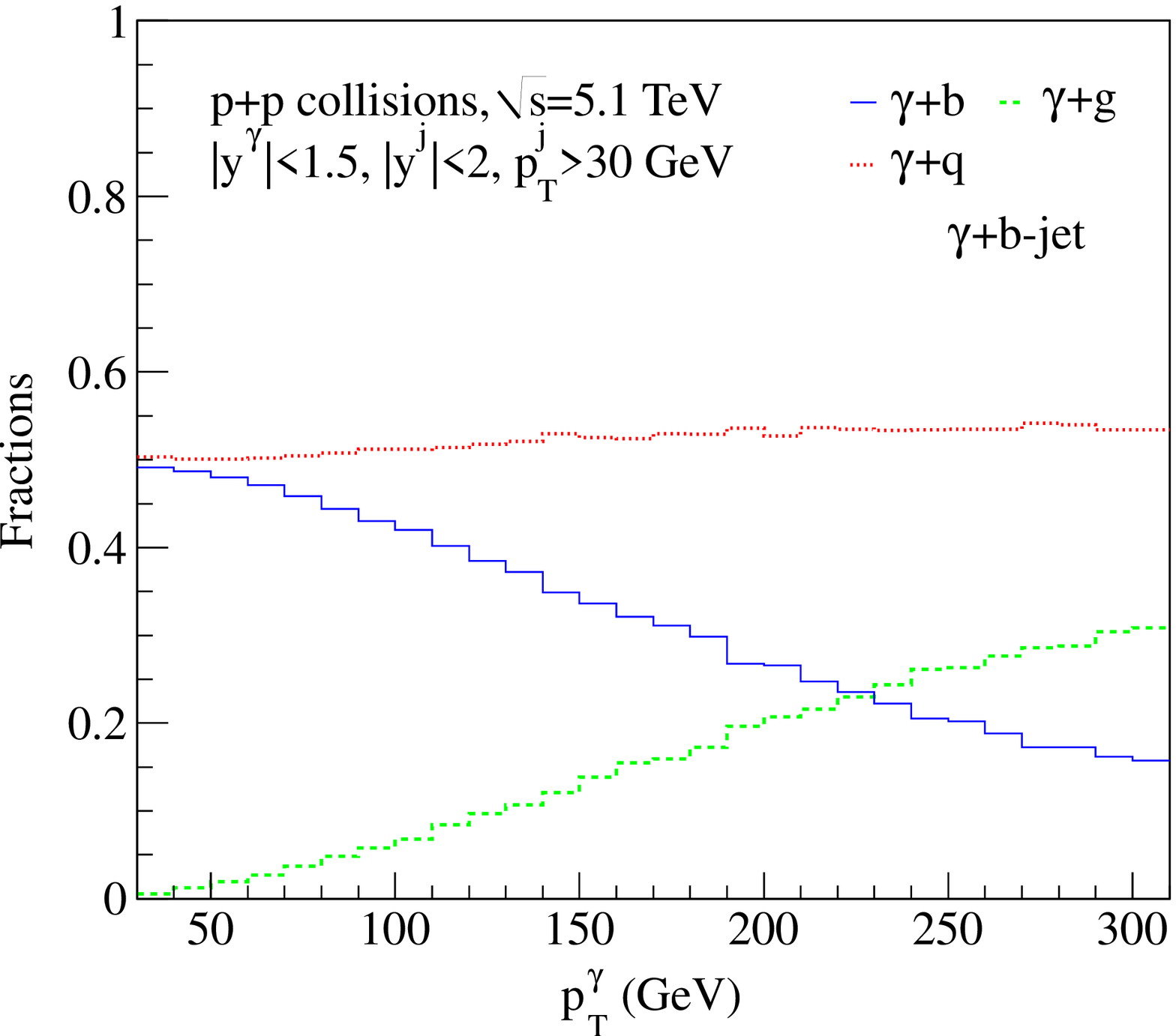, width=2.8in}
\caption{The fractional contributions of different subprocesses to the photon+b-jet production cross section are plotted as a function of the b-jet transverse momentum $p_T^j$ (left) and the photon transverse momentum $p_T^\gamma$ (right) in p+p collisions at $\sqrt{s} = 5.1$~TeV at the LHC. The b-jet is clustered using anti-$k_T$ jet algorithm with $R=0.3$. We have integrated over the b-jet (photon) rapidity $|y^j| < 2$ ($|y^\gamma| < 1.5$), and $p_T^{\gamma, j} > 30$ GeV.}
\label{fig:ratio-bjet}
\eef
\begin{figure}[!b]
\psfig{file=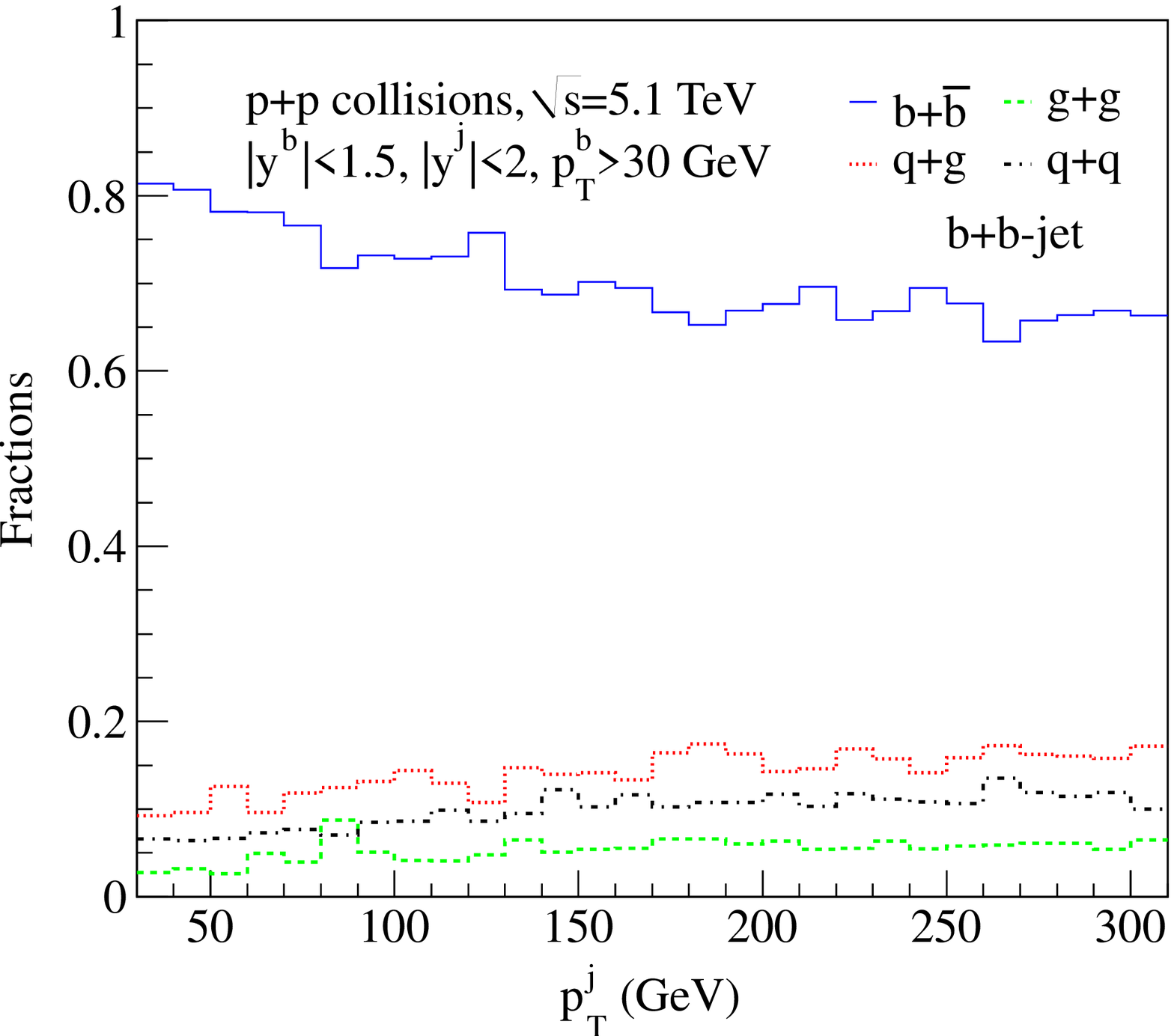, width=2.8in}
\hskip 0.2in
\psfig{file=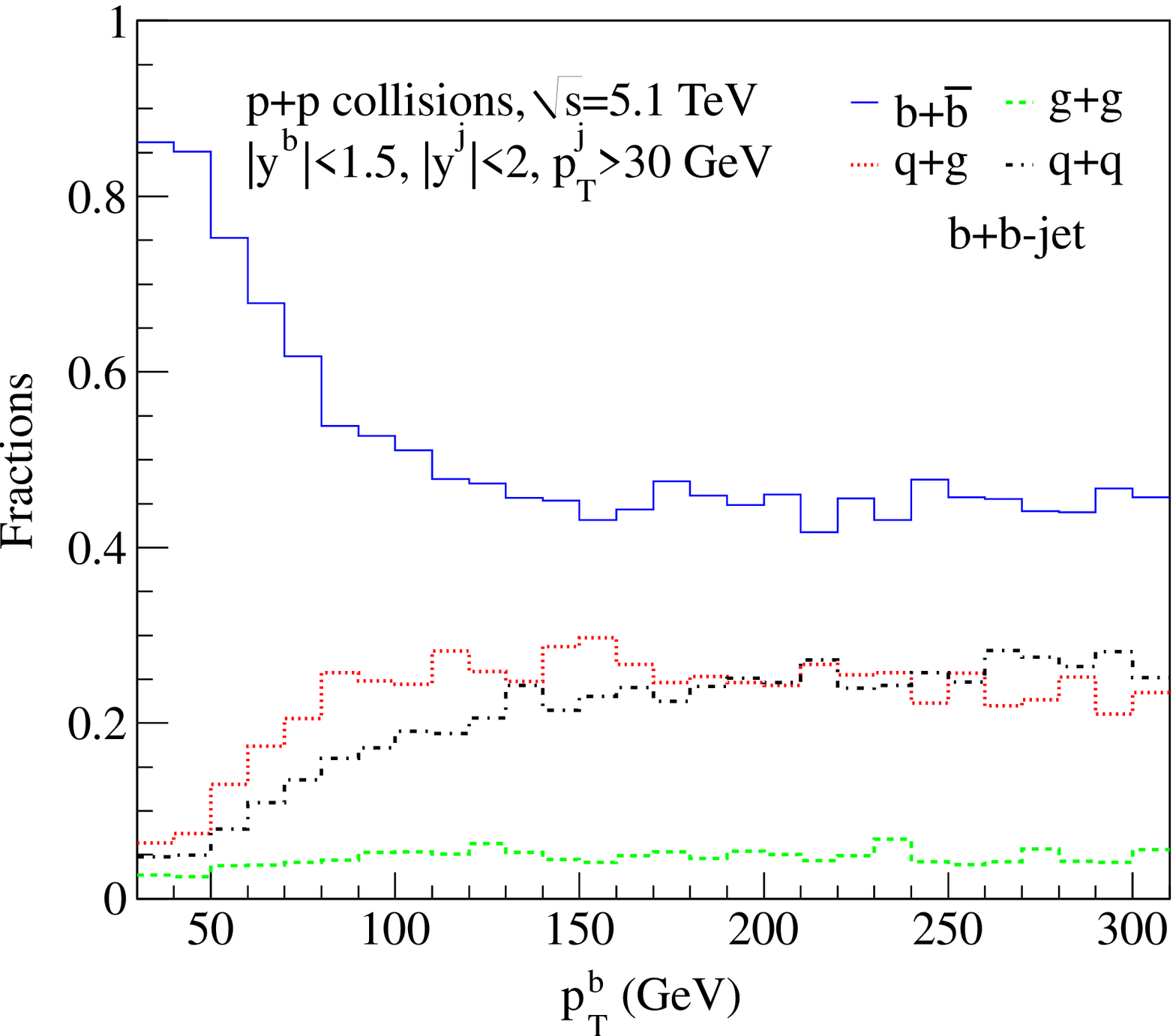, width=2.8in}
\caption{The fractional contributions of different subprocesses to the b-quark+b-jet production cross section are plotted as a function of b-jet transverse momentum $p_T^j$ (left) and b-quark transverse momentum $p_T^b$ (right) in p+p collisions at $\sqrt{s} = 5.1$ TeV at the LHC. The b-jet is clustered using anti-$k_T$ jet algorithm with $R=0.3$. We have integrated over the b-jet (b-quark) rapidity $|y^j| < 2$ ($|y^b| < 1.5$), and $p_T^{b, j} > 30$ GeV.}
\label{fig:ratio-B}
\eef

In order to study energy loss effects and the medium-induced parton shower in heavy ion collisions, we further need detailed information on the flavor origin of b-jets produced in the p+p collisions at $\sqrt{s} = 5.1$ TeV. Heavy quark production in hadronic collisions arises from various elementary $2\to 2$ hard partonic scattering channels, which can be easily separated in the Pythia simulations. In isolated-photon+b-jet production, for example, we have $bg\to \gamma b$, $qg\to \gamma q$, and $q\bar q\to \gamma g$. In the first subprocess, the produced b-quark initiates the b-jet. In heavy ion collisions, the medium modification of such  b-jet production has a direct connection to the physical heavy quark energy loss (mass $m_b$). On the other hand, the latter two cases can also generate b-jets through the parton shower. For the b-jets initiated through $qg\to \gamma q$ or $q\bar q\to \gamma g$, the medium modification would resemble that of a massive quark (color-triplet state) or that of a massive gluon (color-octet state) of effective mass $2m_b$, since a $b\bar{b}$ pair is produced in the shower. We will use subscript symbol ``(s)'' to represent these different final states in the next section. Let us denote their relative contributions as $R_b$, $R_q$, and $R_g$, respectively, and plot them as a function of $p_T^j$ (left) and $p_T^\gamma$ (right) in Fig.~\ref{fig:ratio-bjet}. As one can see from the figure, the $\gamma+b$ contributes  about 50\%  of the cross section when the jet (and photon) transverse momentum are not very large $p_T^{j,\gamma} \lesssim 100$ GeV, and the contribution $R_b$ gradually decreases as these momenta increase. On the other hand, the contribution from $\gamma+g$  gradually increases while the $\gamma+q$ contribution remains  constant as $p_T^{j,\gamma}$ increases.

In this study, we focus on energetic boosted B-mesons, where the fragmentation process occurs predominantly outside of the medium  and 
the interactions in QGP that we will consider happen at the partonic level~\cite{Markert:2008jc,Sharma:2009hn}.
For B-meson-tagged b-jet production we first simulate the b-quark+b-jet cross section in Pythia 8. Once we implement all the interactions with the dense medium, we  apply the $b\to B$ meson fragmentation function to convert the b-quark to a B-meson. At the partonic level, b-quark+b-jet production arises from: (1) $q\bar q\to b\bar b$, $gg\to b\bar b$; (2)  $gg\to gg$, $q\bar q\to gg$; (3) $qg\to qg$; (4) $qq\to qq$, $gg\to q\bar q$, $q\bar q\to q\bar q$. For the first case (1), we have $b\bar b$ in the final states, thus one b-quark initiates the b-jet, while the other b-quark fragments into the B-meson. This subprocess represents, on both the near side and the away side,
 the energy loss of a b-quark. For the remaining cases, the b-quark is produced in the parton shower and, thus, will mimic the energy loss of a massive gluon or a massive quark (see above). We implement the energy loss for such states in the formalism accordingly. The relative contributions of the various channels are presented in Fig.~\ref{fig:ratio-B} as a function of $p_T^j$ (left) and $p_T^b$ (right), respectively. As one can see  from this figure,  the $b+\bar b$ contributions dominate in the whole kinematic region, which implies that the B-meson-tagged b-jet production has a much more direct connection  to prompt b-quark energy loss.

\section{Modification of tagged b-jet production in dense QCD matter}
\label{AA}
In this section, we present the medium-modification of  $\gamma$-tagged and B-tagged b-jet production in the ambiance of the QGP. Such final-state hot medium effects include radiative energy loss, caused by medium-induced parton splitting, as well as the dissipation of the energy of the parton shower through collisional interactions in the strongly interacting plasma~\cite{Neufeld:2011yh,Neufeld:2014yaa}. The medium-modified $\gamma$+b-jet cross section per binary nucleon-nucleon scattering is calculated as follows~\cite{Neufeld:2012df,Dai:2012am,Huang:2013vaa}: 
\bea
\frac{1}{\langle N_{\rm bin}\rangle} \frac{d\sigma^{AA}}{dp_T^\gamma dp_T^j} = 
\sum_{(s)} \int_0^1 d\epsilon P_{(s)}(\epsilon) J_{(s)}(\epsilon) 
\frac{d\sigma_{(s)}^{\rm LO+PS}\left(p_T^\gamma, J_{(s)}(\epsilon) p_T^j\right)}{dp_T^\gamma dp_T^j},
\eea
where $\langle N_{\rm bin}\rangle$ is the average number of binary nucleon-nucleon interactions,  the subscript ``(s)'' represents different final partonic states as described in Sec.~\ref{pp}, and $P_{(s)}(\epsilon)$ is the probability distribution that a fraction $\epsilon$ of the hard scattering parton energy is converted to a medium-induced parton shower for a system with color state ``(s)''. The hard production points are distributed according to the binary nucleon-nucleon collision density and the jets propagate through the medium that follows the participant number density and undergoes Bjorken expansion, which are determined by an optical Glauber model with an inelastic nucleon-nucleon cross section $\sigma_{in} = 70$ mb at $\sqrt{s} = 5.1$ TeV~\cite{d'Enterria:2003qs}. On the other hand, $J_{(s)}(\epsilon) \equiv J_{(s)}(\epsilon; R, \omega^{\rm coll})$ is a phase space Jacobian
\bea
J_{(s)}(\epsilon; R, \omega^{\rm coll}) =  \left\{ 1- \left[1- f_{(s)}(R, \omega^{\rm coll})\right] \epsilon  \right\}^{-1},
\eea
where $f_{(s)}(R, \omega^{\rm coll})$ is the fraction of the medium-induced parton shower energy that is retained inside the jet cone of radius $R$, as opposed to ``lost'' outside. Such a fraction is evaluated from the medium-induced gluon distribution $\frac{d^2N^g}{d\omega dr}$ ($\omega$ is the energy and $r$ is the angle)~\cite{Vitev:2008rz,Huang:2013vaa}:
\bea
f_{(s)}(R, \omega^{\rm coll}) = \left( \int_0^R dr \int_{\omega^{\rm coll}}^E d\omega \frac{\omega d^2N^g_{(s)}}{d\omega dr} \right) \Big/
\left( \int_0^{R\infty} dr \int_{0}^E d\omega \frac{\omega d^2N^g_{(s)}}{d\omega dr} \right),
\eea
where $\omega^{\rm coll}$  simulates the effects of collisional energy loss discussed in~\cite{Neufeld:2011yh,Huang:2013vaa}. It is obtained as a solution 
to  
\bea
 \int_0^{R\infty} dr \int_{\omega^{\rm coll}}^E d\omega \, \frac{\omega d^2N^g_{(s)}}{d\omega dr}    =  \sum_{i\in \rm shower} \Delta \epsilon_{i}^{\rm coll}  
\equiv  \Delta E^{\rm coll}_{\rm shower}.
\eea
Here, $\Delta E^{\rm coll}_{\rm shower}$ is part of the energy of the medium-induced parton shower dissipated in the QGP through collisional 
processes that was obtained in~\cite{Neufeld:2011yh}.
To generate the medium-modified B-meson+b-jet cross section per binary collision, we first calculate the medium-modified b-quark+b-jet cross section as follows~\cite{He:2011pd,Huang:2013vaa}
\bea
\frac{1}{\langle N_{\rm bin}\rangle} \frac{d\sigma^{AA}}{dp_T^b dp_T^j} = \sum_{(s,s')} \int_0^1 d\epsilon P_{(s)}(\epsilon) \frac{1}{1-\epsilon} 
\int_0^1 d\epsilon' P_{(s')}(\epsilon') J_{(s')}(\epsilon') 
\frac{d\sigma_{(s,s')}^{\rm LO+PS}\left(p_T^b/(1-\epsilon), J_{(s')}(\epsilon') p_T^j\right)}{dp_T^b dp_T^j},
\label{eq:b+b-jet}
\eea
where $(s)$ and $(s')$ are the final-state parton color states in the hard partonic scattering level in the Pythia simulation, that further transmute into the b-jet and b-quark (this b-quark later on hadronizes into the B-meson), respectively. We then use the $b\to B$ fragmentation function $D_{b\to B}(z)$ to fragment into the desired B-meson-tagged b-jet case:
\bea
\frac{1}{\langle N_{\rm bin}\rangle} \frac{d\sigma^{AA}}{dp_T^B dp_T^j} = \int_0^1 \frac{dz}{z} D_{b\to B}(z)  \frac{1}{\langle N_{\rm bin}\rangle} \frac{d\sigma^{AA}}{dp_T^b dp_T^j}.
\eea
Here, the momentum fraction $z=p_T^B/p_T^b$, where $p_T^B$ is the B-meson transverse momentum, and we use the B-meson fragmentation functions as  in~\cite{Vitev:2006bi,Sharma:2009hn}.

\begin{figure}[!t]
\psfig{file=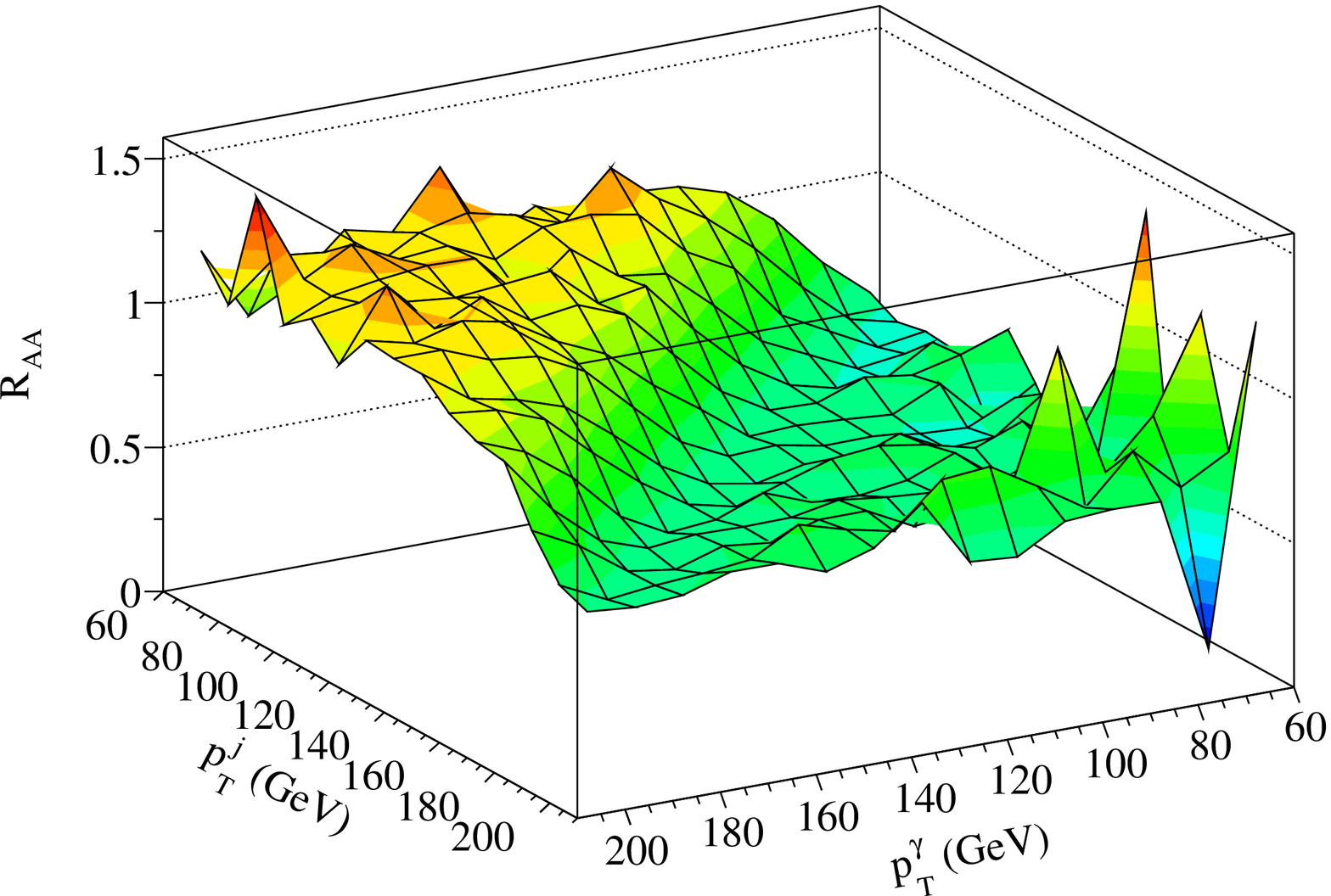, width=3.2in}
\hskip 0.2in
\psfig{file=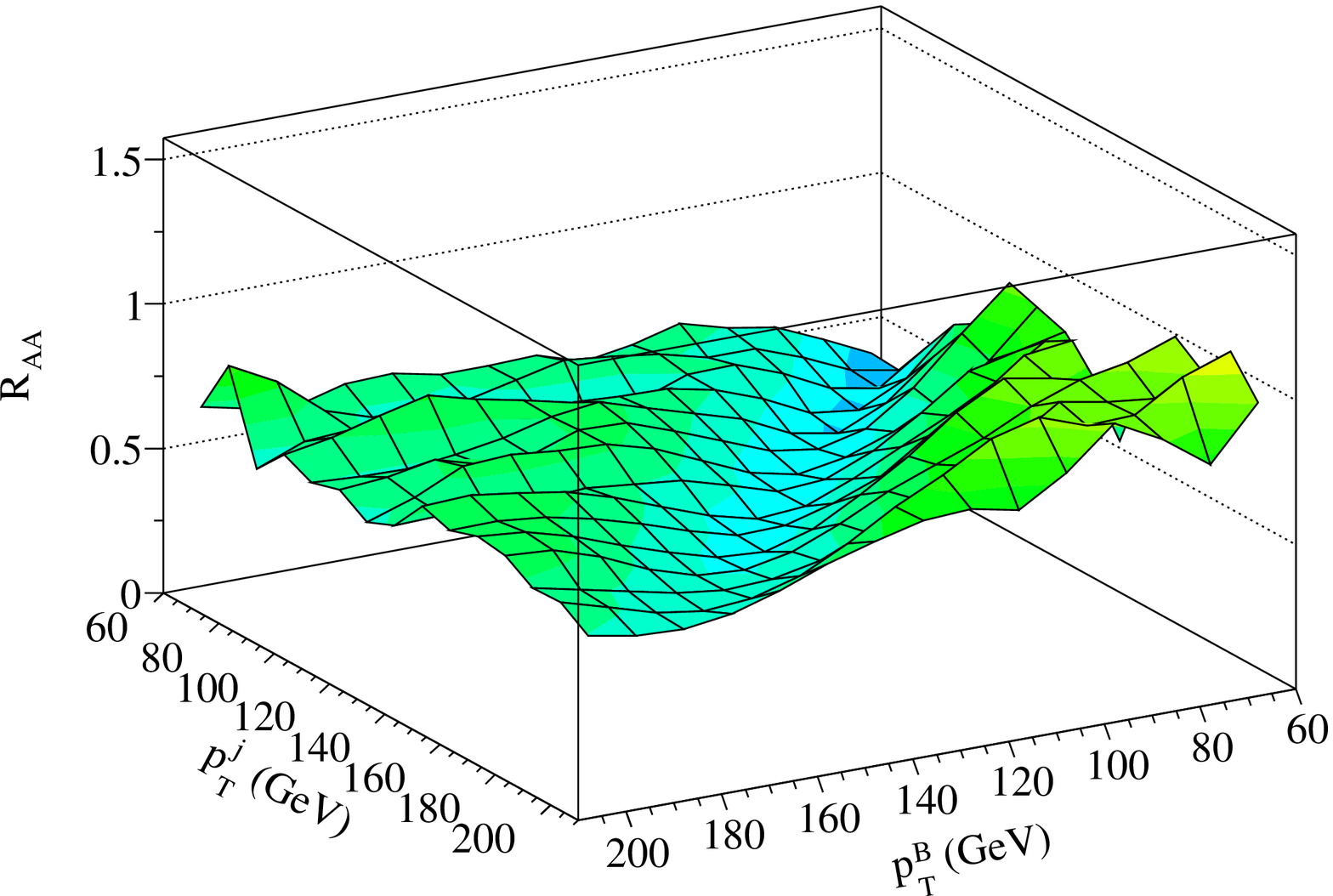, width=3.2in}
\caption{Left: The nuclear modification factor $R_{AA}$ for $\gamma$+b-jet production in Pb+Pb collisions at LHC $\sqrt{s} = 5.1$ TeV is plotted as a function of the b-jet transverse momentum $p_T^j$ and the photon $p_T^\gamma$. Rapidities are integrated over $|y^{\gamma}|<1.5$ and $|y^{j}| < 2$. Right: $R_{AA}$ for B-meson-tagged b-jet production as functions of b-jet $p_T^j$ and B-meson $p_T^B$. The rapidity cuts for B-meson and b-jet are $|y^B|<1.5$ and $|y^{j}| < 2$. The coupling between the jet and the medium is $g_{\rm med} = 2.0$, and the mass for the collimated propagating of parent parton system for the b-jet is set to be $M=2m_b$. In these figures the suppression ($R_{AA} <1$) and enhancement ($R_{AA} <1$) are color coded, which can be seen in the online
color version.}
\label{fig:RAA}
\eef

Let us now turn to the phenomenological results for  tagged b-jet production in Pb+Pb collisions at the LHC. 
Our simulations are performed at a center-of-mass energy per nucleon-nucleon pair $\sqrt{s_{NN}} = 5.1$ TeV for direct comparison to the planned experimental measurements at the LHC. We focus on the most central collisions with average number of participants $N_{\rm part} = 360$ and the central rapidity region with photon or B-meson rapidity $|y^{\gamma, B}| < 1.5$ and b-jet rapidity $|y^j| < 2$. We choose the b-jet reconstruction radius $R=0.3$ and require back-to-back configuration $|\phi_{\gamma, B} - \phi_j| > 3/4\pi$. We implement the isolation cut for the photon that requires the hadronic transverse energy in the cone of radius $R_{\rm iso} = 0.4$ around the photon direction to be less than 7\% of the photon energy. 

The many-body QCD dynamics that modifies the tagged b-jet production in heavy ion collisions will be manifest in the deviation of the A+A cross section  from the binary collision scaled p+p results, 
\bea
R_{AA}^{\gamma+b\text{-}jet} = \left(\frac{d\sigma^{AA}}{dp_T^\gamma dp_T^j}\right) \Big/  \left( \langle N_{\rm bin}\rangle \frac{d\sigma^{pp}}{dp_T^\gamma dp_T^j} \right),
\qquad
R_{AA}^{B+b\text{-}jet} = \left( \frac{d\sigma^{AA}}{dp_T^B dp_T^j}\right)   \Big/  \left( \langle N_{\rm bin}\rangle \frac{d\sigma^{pp}}{dp_T^B dp_T^j} \right).
\eea
Theoretical predictions for the nuclear modification factor $R_{AA}$ are presented in Fig.~\ref{fig:RAA}, for $\gamma$+b-jet production (left) and B-meson+b-jet production (right), respectively. We include both collisional and radiative energy loss effects, and the mass $M$ of the propagating parton system in the final state is implemented as  in~\cite{Huang:2013vaa}. In this illustration for the 3D plots of $R_{AA}$, we set $M=2m_b$, twice the b-quark mass, and choose the coupling between the jet and the medium $g_{\rm med} = 2.0$~\cite{Kang:2014xsa,Huang:2013vaa}. The coupling between the final-state partons and the medium is the largest uncertainty in our nuclear modification model. It affects the absolute value of the cross section suppression/enhancement but does not alter the qualitative behavior of $R_{AA}$, which we now discuss.  As one can see clearly,  the largest suppression is observed when the transverse momenta of the trigger particle and the b-jet are similar: $p_T^{\gamma,B} \approx p_T^j$. For $p_T^j < p_T^\gamma$, there can be an enhancement of the cross section for $\gamma$+b-jet production, which is consistent with what was observed in $\gamma$+light-flavor-jet production~\cite{Chatrchyan:2012gt,Dai:2012am}. For $p_T^j > p_T^\gamma$, the b-jet suppression is somewhat smaller than the quenching of inclusive b-jets~\cite{Huang:2013vaa,Chatrchyan:2013exa}, because the fraction of $\gamma$-tagged b-jets originating from a prompt gluon is smaller. The fluctuations arise from the Pythia 8 limited statistics for small, steeply falling baseline cross sections, especially when $p_T^\gamma$ and $p_T^j$ are significantly different from each other. 
On the other hand, we find that the overall suppression is stronger in B-meson+b-jet production than that of photon+b-jet production since both the parton that fragments into the B-meson and the b-jet lose energy when they traverse the medium, while the isolated-photon does not lose energy. When $p_T^j \gg p_T^B$, or $p_T^j \ll p_T^B$  there can be a strong enhancement in $R_{AA}$~\cite{He:2011pd}, which we don't see in the more limited kinematic range that we study.  The overall 2D suppression pattern is more uniform than for light jets.

\begin{figure}[!t]
\psfig{file=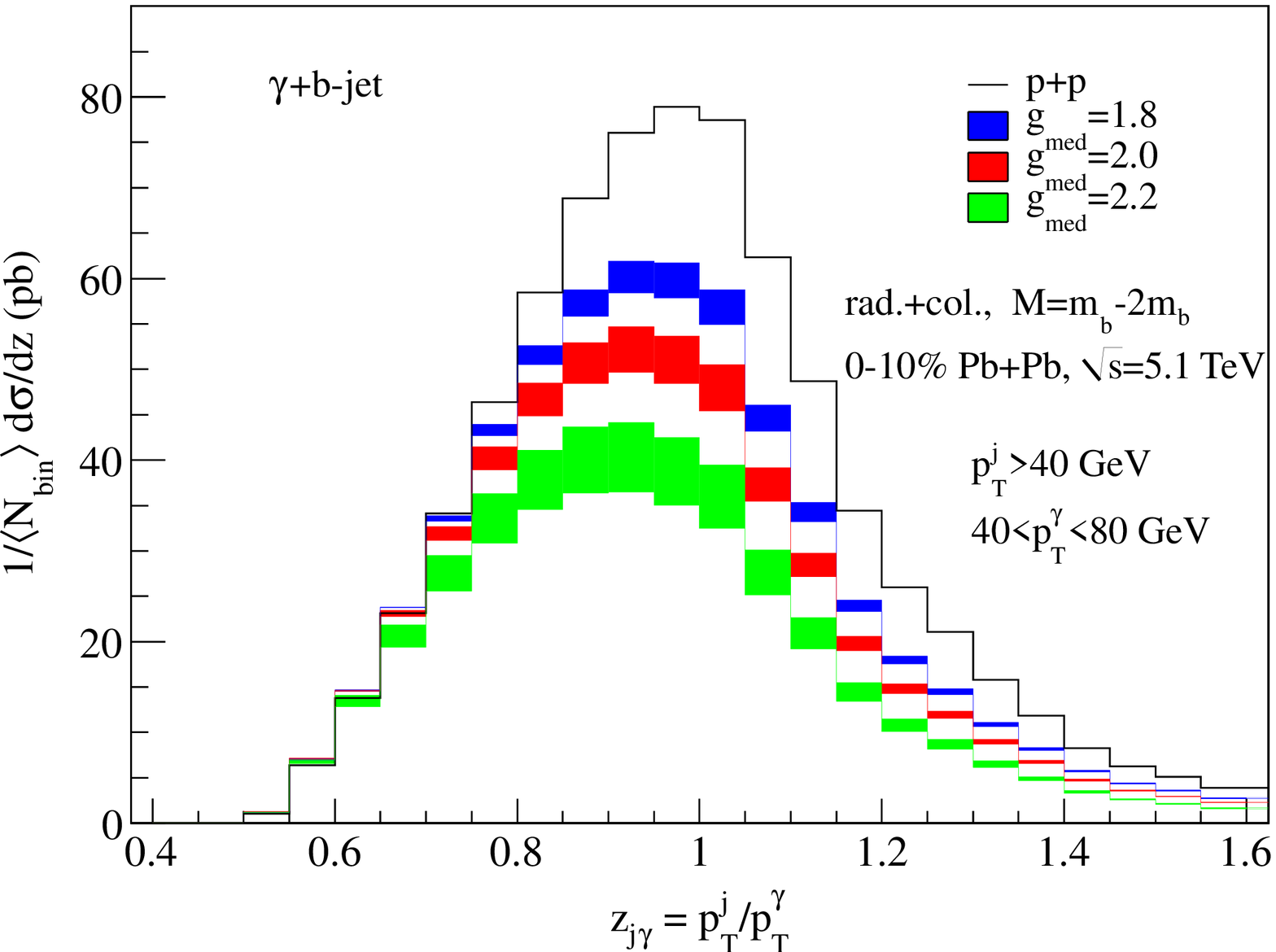, width=3.2in}
\hskip 0.2in
\psfig{file=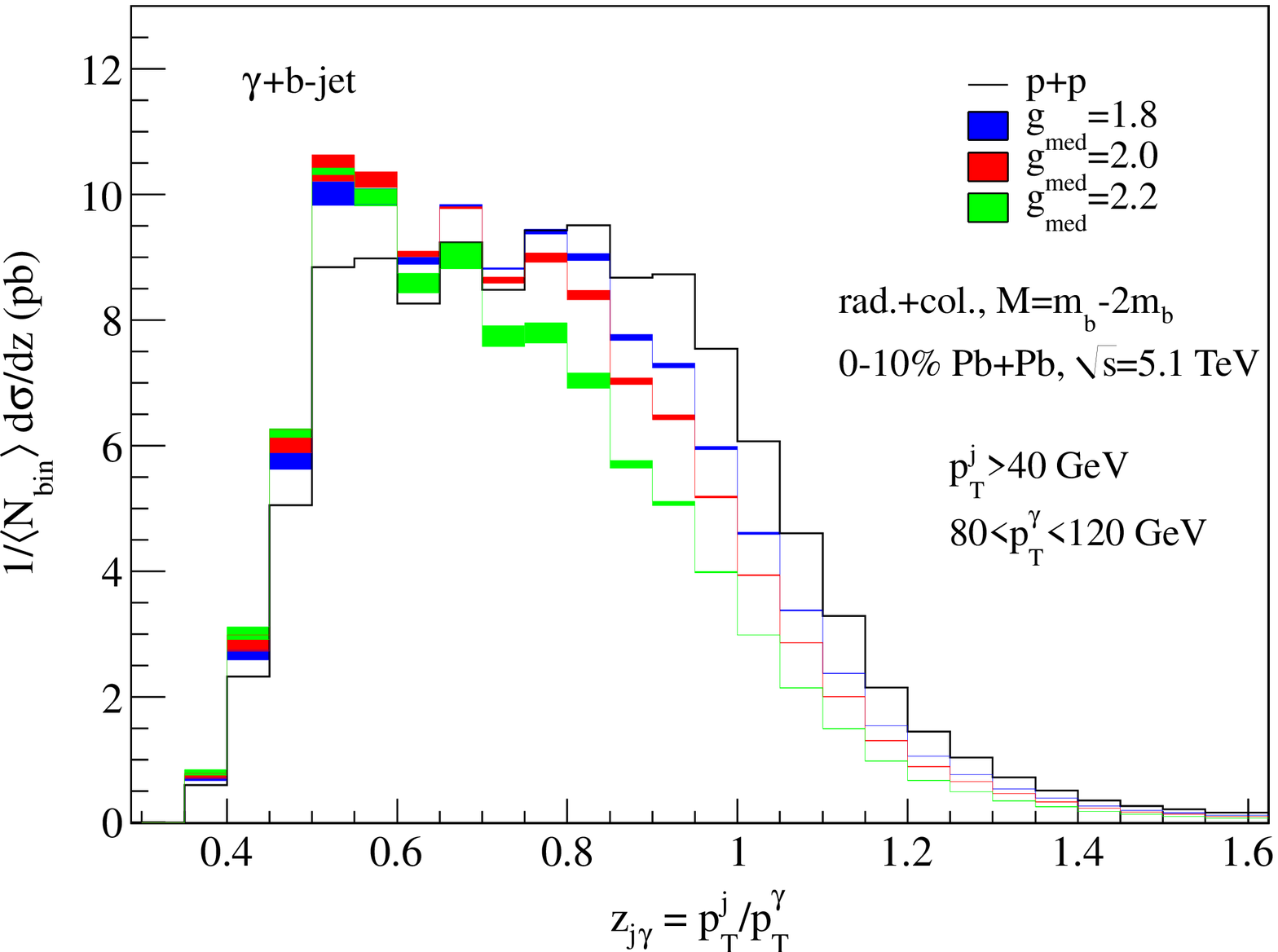, width=3.2in}
\caption{The isolated-photon-tagged b-jet asymmetry distribution is shown for different coupling strengths between the jet and the medium. We have b-jet transverse momentum $p_T^j > 40$ GeV. Left panel: $40<p_T^\gamma<80$ GeV. Right Panel: $80 < p_T^\gamma < 120$ GeV. Bands correspond to a range of masses of the propagating system between $m_b$ and $2m_b$. Blue, red, and green bands correspond to $g_{\rm med} = 1.8, 2.0, 2.2$, respectively.}
\label{fig:z-photon}
\eef
\begin{figure}[!t]
\psfig{file=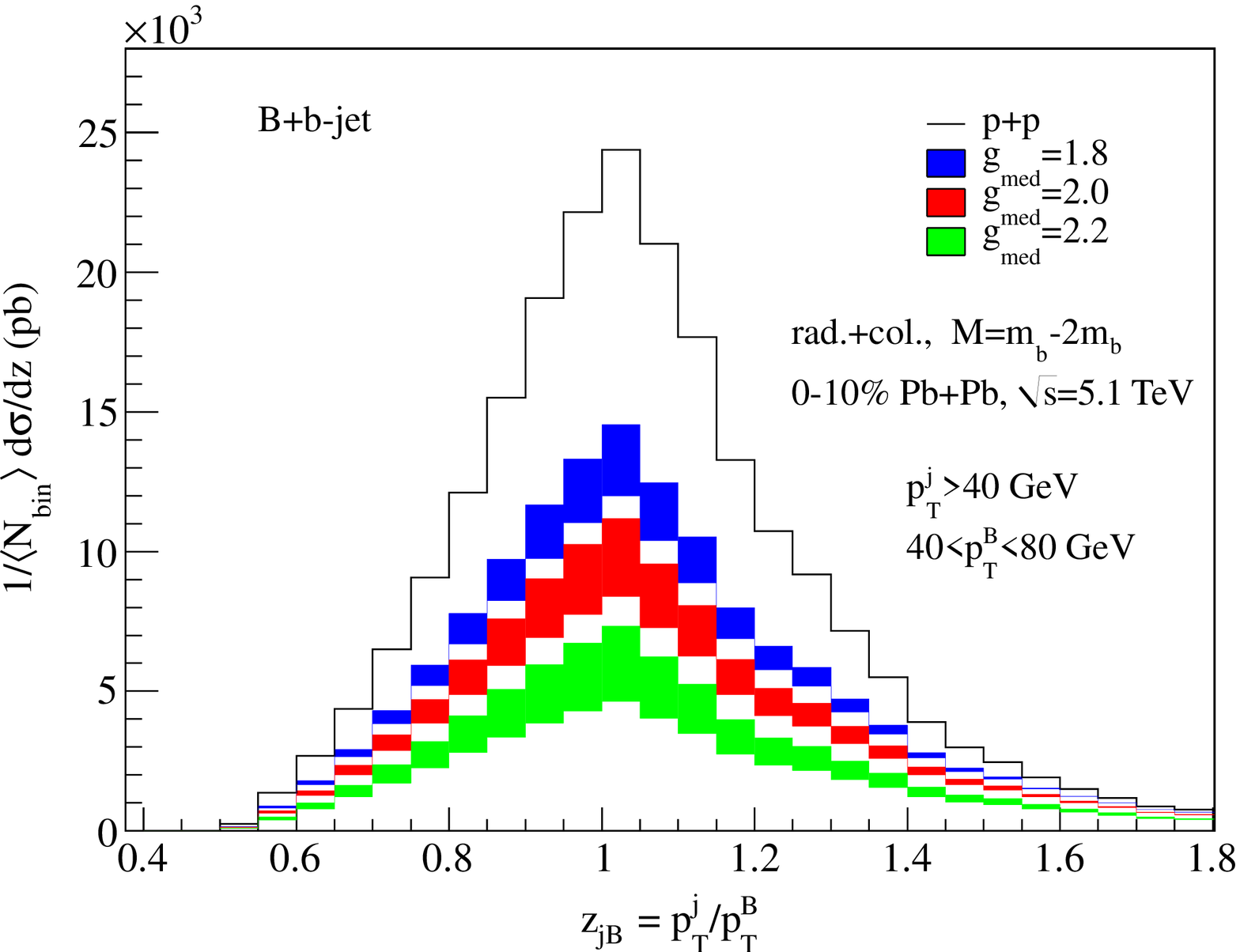, width=3.2in}
\hskip 0.2in
\psfig{file=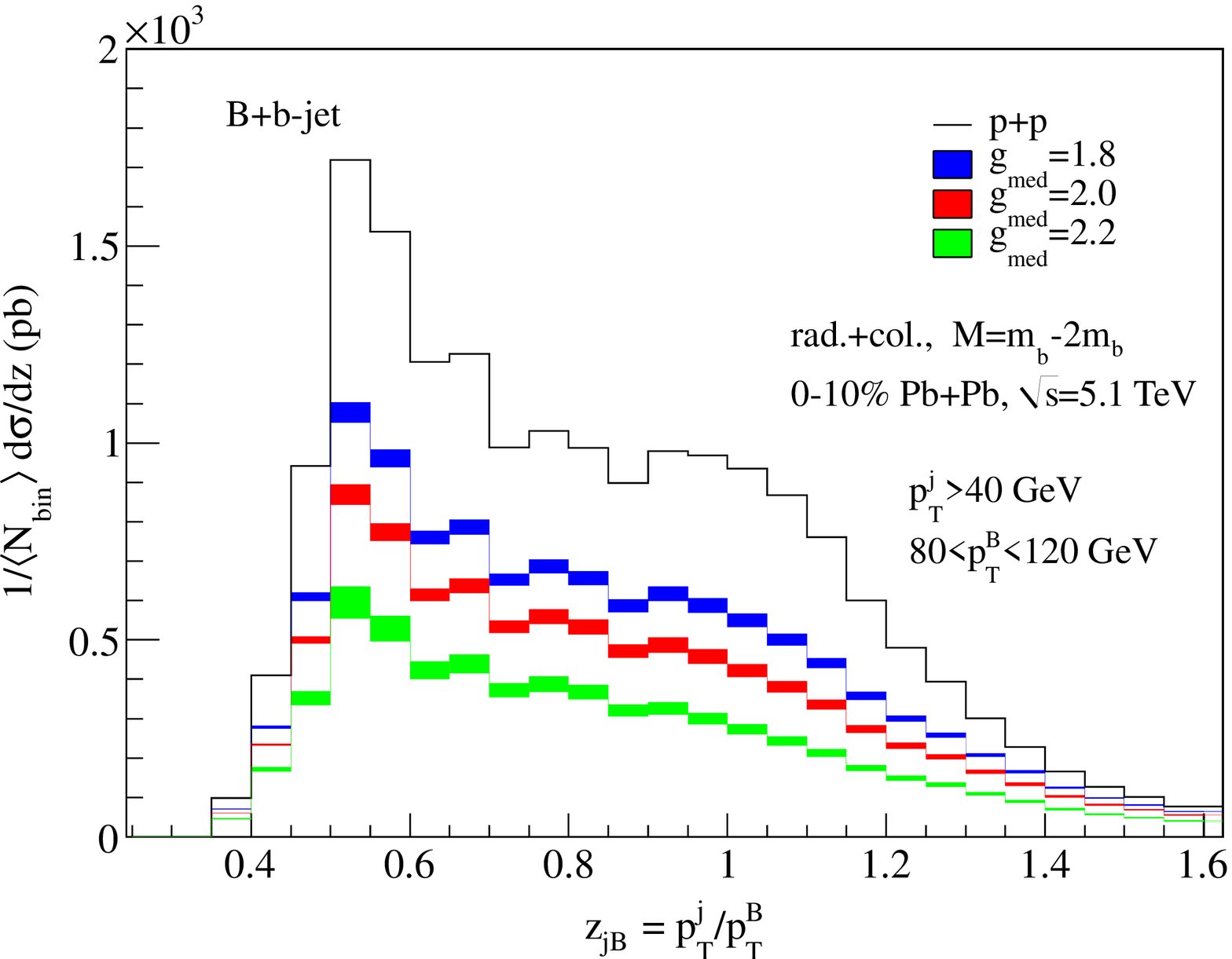, width=3.2in}
\caption{The B-meson-tagged b-jet asymmetry distribution is shown for different coupling strengths between the jet and the medium. 
Kinematic cuts and jet-medium couplings are the same as in Fig.~\ref{fig:z-photon} with the replacement $\gamma \rightarrow$B.}
\label{fig:z-B}
\eef

Next, we consider the tagged b-jet event asymmetry. For this purpose, we define the following variables to represent the momentum imbalance of the $\gamma$+b-jet as well as B+b-jet events, 
$ z_{j\gamma} = {p_T^j}/{p_T^\gamma}$, $z_{jB} = {p_T^j}/{p_T^B}$.
Based on our simulations of the double differential cross sections $d\sigma/dp_T^{\gamma, B} dp_T^j$ above, we 
evaluate the $z$-distributions as follows:
\bea
& \frac{d\sigma}{dz_{j\gamma}} = \int_{p_T^{j, \rm min}}^{p_T^{j, \rm max}} dp_T^j\, \frac{p_T^j}{z_{j\gamma}^2} \frac{d\sigma(p_T^\gamma = p_T^j/z_{j\gamma})}{dp_T^\gamma dp_T^j},  \quad
\frac{d\sigma}{dz_{jB}} = \int_{p_T^{j, \rm min}}^{p_T^{j, \rm max}} dp_T^j\, \frac{p_T^j}{z_{jB}^2} \frac{d\sigma(p_T^B = p_T^j/z_{jB})}{dp_T^B dp_T^j},
\eea
where $p_T^{j, \rm min}$ and $p_T^{j, \rm max}$ are specified by the experimental cuts on the b-jet transverse momentum. In our predictions for the future Pb+Pb run at the LHC we choose the b-jet transverse momentum $p_T^{j} > 40$ GeV. On the other hand, we explore two different transverse momentum cuts for the $\gamma$ and  the B-meson: (1) $40 < p_T^{\gamma, B} < 80$ GeV, (2) $80 < p_T^{\gamma, B} < 120$ GeV. We see that  cut (2) leads to more asymmetric distribution of $\gamma$(B)+b-jet production than  cut (1). 

The momentum imbalance distributions are given in Fig.~\ref{fig:z-photon} for $\gamma$+b-jet production, and in Fig.~\ref{fig:z-B} for B+b-jet production, respectively. Note 
that here we present the absolute differential cross section $d\sigma/dz$ instead of the usually measured normalized asymmetry distribution $\left(1/\sigma\right)d\sigma/dz$~\cite{He:2011pd,Dai:2012am,Chatrchyan:2012gt,Chatrchyan:2011sx,Aad:2010bu}. In both plots the left panel corresponds to cut selection (1), while the right panel corresponds to cut (2) -- the more asymmetric one. The black histogram corresponds to the p+p baseline. The colored histograms correspond to the cross sections per binary collision in central 0-10\% Pb+Pb collisions, with bands representing a range of masses of the propagating system between $m_b$ and $2m_b$. Blue, red, and green bands correspond to $g_{\rm med} = 1.8, 2.0, 2.2$, respectively. The range of these values has worked well in describing  and/or  predicting the dijet~\cite{He:2011pd} and photon-tagged jet~\cite{Dai:2012am} asymmetry distribution, as well as inclusive b-jet~\cite{Huang:2013vaa} and hadron~\cite{Kang:2014xsa} suppression at the LHC. 

\begin{table}
\caption{Theoretical results for $\langle z_{j\gamma}\rangle$ ($\langle z_{jB}\rangle$) in photon-tagged (B-tagged) b-jet production in p+p and central Pb+Pb reactions at center-of-mass energy $\sqrt{s_{NN}} = 5.1$ TeV at LHC. For the photon (B-meson), we have two kinematic cuts, while $p_T^j > 40$ GeV for the b-jet.}
\label{table:photon-B}
\begin{center}
\begin{tabular}{l | c c | c c }
\hline
\hline
\multirow{2}{*}{System}
&\multicolumn{2}{ c| }{$\langle z_{j\gamma} \rangle$} & \multicolumn{2}{ c }{$\langle z_{jB} \rangle$} \\
\cline{2-5}
  &  \quad  $40<p_T^\gamma<80$ GeV & \quad $80<p_T^\gamma<120$ GeV 
  &  \quad  $40<p_T^B<80$ GeV & \quad $80<p_T^B<120$ GeV   \\ 
\hline
p+p & 0.98 & 0.78  & 1.04 & 0.82 \\
A+A, $g = 1.8$, $M=2m_b$ & 0.95 & 0.74 & 1.06 & 0.83 \\
A+A, $g = 1.8$, $M=m_b$ & 0.95 & 0.74 & 1.07 & 0.83 \\
A+A, $g = 2.0$, $M=2m_b$ & 0.94 & 0.73 & 1.07 & 0.83 \\
A+A, $g = 2.0$, $M=m_b$ & 0.94 & 0.73 & 1.09 & 0.83 \\
A+A, $g = 2.2$, $M=2m_b$ & 0.93 & 0.71 & 1.08 & 0.83 \\
A+A, $g = 2.2$, $M=m_b$ & 0.93 & 0.71 & 1.10 & 0.83 \\
\hline
\hline
\end{tabular}
\end{center}
\end{table}

For the $\gamma$+b-jet case in Fig.~\ref{fig:z-photon} and the more symmetric cut (1) we see a moderate suppression of the cross section which is strongest when $p_T^j \sim p_T^\gamma$ and  $z_{j\gamma}\sim 1$. At the same time, the final-state interactions in the dense medium will shift the asymmetry variable $z_{j\gamma}$ down to smaller values. This arises from the energy loss of the b-jet, while the photon escapes out of the medium unscathed.   On the other hand, for the more asymmetric cut (2), we find an enhancement for $z_{j\gamma} \sim 0.6$ and a change in the shape of the imbalance distribution. These features are all consistent with the asymmetric suppression pattern seen in the 3D plot of  $R_{AA}$ in Fig.~\ref{fig:RAA}. Note that the fluctuation/uncertainty in $R_{AA}$ should only have minimal influence in our plot here since they happen mainly in the region when $p_T^\gamma$ and $p_T^j$ are significantly different from each other (by a factor of 3 or more), which corresponds to $z_{j\gamma}$ blow $1/3$ or above $3$ and has much smaller cross sections. To further quantify such  down-shift in $z_{j\gamma}$ distribution~\footnote{Here and in the rest of the paper, down-shift or up-shift are always relative to the p+p baseline.}, we define the mean value of $z_{j\gamma}$ (similarly also for $z_{jB}$) as
\bea
\langle z_{j\gamma}\rangle =\left( \int dz_{j\gamma} z_{j\gamma} \frac{d\sigma}{dz_{j\gamma}} \right)
\Big/  \left( \int dz_{j\gamma} \frac{d\sigma}{dz_{j\gamma}} \right) ,
\qquad
\langle z_{jB}\rangle =\left( \int dz_{jB} z_{jB} \frac{d\sigma}{dz_{jB}} \right) \Big/ 
\left( \int dz_{jB} \frac{d\sigma}{dz_{jB}} \right),
\eea
and show its value $\langle z_{j\gamma}\rangle$ in Table.~\ref{table:photon-B} for $\gamma$+b-jet production. From these calculations, we see there is around 5\% (10\%) down-shift in $\langle z_{j\gamma}\rangle$ for the more symmetric cut (1) (more asymmetric cut (2)), which is somewhat smaller than those values ($\sim 15\%$) in the $\gamma$+light-jet production, as studied in~\cite{Dai:2012am} and measured in~\cite{Chatrchyan:2012gt}.

On the other hand, for the B+b-jet case in Fig.~\ref{fig:z-B} we find strong suppression for both cuts, (1) and (2). For the more symmetric cut (1), as can be seen in Fig.~\ref{fig:z-B} (left), the distribution is peaked slightly above $z_{jB}=1$. This is also illustrated by the average value $\langle z_{jB}\rangle$ shown in Table~\ref{table:photon-B}. The reason for such behavior is that it is the $p_T^b$  
of the b-quark (bigger than the $p_T^B$  of the B-meson) that balances the transverse momentum of the away-side jet.  Besides suppression,  in contrast to $\gamma$+b-jet production there is no noticeable down-shift of the asymmetry  $z_{jB}$ distribution. The calculated $\langle z_{jB}\rangle$ for both cuts even show a slight increase in going  from p+p to Pb+Pb collisions, as seen from Table~\ref{table:photon-B}. This can be understood as follows: as we discussed in the previous section, the $b+\bar{b}$ final state dominates  b+b-jet production, recall Fig.~\ref{fig:ratio-B}. If we consider this dominant  $b+\bar{b}$ channel and inspect Eq.~\eqref{eq:b+b-jet}, on average the b-jet  loses less energy,  
$(1-f_{(s)}(R, \omega^{\rm coll})) \epsilon$, than the b quark, $\epsilon$. This, in turn, leads to 
 a slight up-shift of $z_{jB}$ to larger values. 
This behavior is very different from the  light dijet asymmetry enhancement observed in  experiments~\cite{Chatrchyan:2011sx,Aad:2010bu}, which corresponds to a momentum imbalance
 down-shift. It suggests that the large asymmetry enhancement for light dijets is likely driven 
by the different energy loss for quarks and gluons, proportional to $C_F$ and $C_A$, in the dominant
q+g final-state channel. The value of the up-shift reported in Table~\ref{table:photon-B}  might have a small numerical dependence on the details of fragmentation into B-mesons, but this will not change the qualitative difference in the 
behavior of $\langle z_{j\gamma}\rangle$ and $\langle z_{jB}\rangle$.    It will be important in the  future to use  tagged b-jet  Pb+Pb measurements at the LHC to elucidate the precise physics mechanisms responsible for 
the large nuclear modification of heavy flavor production in heavy ion collisions.

\section{Conclusions}
In summary, we presented theoretical predictions for the 2D nuclear modification $R_{AA}$ and the related momentum imbalance shift
of isolated-photon-tagged  and B-meson-tagged b-jets in Pb+Pb collisions at $\sqrt{s_{NN}} = 5.1$ TeV at the LHC.
We validated the Pythia 8 simulations of tagged b-jet cross sections in nucleon-nucleon collisions through comparison of photon-tagged b-jet results
with p+${\rm \bar p}$ data at $\sqrt{s} = 1.96$ TeV from the Tevatron. We also found that particle tagging can significantly increase the fraction of recoiling b-jets that
originate from prompt b-quark relative to the inclusive b-jet case. While in the latter case this fraction can be as low as 20\%, B-meson tagging in particular can increase
the contribution of prompt b-quarks to 70-80\%. In Pb+Pb collisions in the LHC we further considered the medium-induced parton shower in the
soft gluon energy loss limit and any additional dissipation of its energy in the QGP due to collisional interactions.
We found significant nuclear suppression when the trigger particle momentum is similar to the b-jet momentum $p_T^{\gamma, B}\sim p_T^j$.
A modest $5-10\%$ increase in the transverse momentum imbalance for $\gamma$+b-jet production in Pb+Pb collisions is predicted from our calculations,
depending on the specific kinematic cuts, which is slightly smaller than that observed for $\gamma$+light jet production.
For B-tagged b-jets we found an even larger suppression of the double differential cross section since both the jet and the tagging b-quark lose energy.
On the other hand, the asymmetry variable $z_{jB}$ in B+b-jet production shows a slight increase from p+p to A+A collisions, a behavior quite different from
the one observed in dijet asymmetry distributions. By using the flexibility of b-jet tagging, comparison of upcoming experimental measurements to theoretical
calculations, such as the ones presented here, can provide us with unique new insights into heavy flavor dynamics in the strongly-interacting plasma.
In the future, we expect that the theoretical uncertainty of b-jet production in nuclear matter can be further reduced by
going beyond the current radiative energy loss approximations~\cite{Ovanesyan:2011kn,Kang:2014xsa} and by including  higher order computations
for multiple scattering~\cite{Kang:2013raa,Kang:2014ela} with full account of heavy quark mass effects.

\section*{Acknowledgments}
This research is supported by the US Department of Energy, Office of 
Science under  Contract No. DE-AC52-06NA25396, by the DOE Early Career Program”, and in 
part by the LDRD program at Los Alamos National Laboratory.

\bibliographystyle{h-physrev5}   
\bibliography{biblio}

\begin{thebibliography}{10}

\bibitem{Owens:1986mp}
J.~Owens,
\newblock Rev.Mod.Phys. {\bf 59}, 465 (1987).

\bibitem{Sterman:1977wj}
G.~F. Sterman and S.~Weinberg,
\newblock Phys.Rev.Lett. {\bf 39}, 1436 (1977).

\bibitem{Wang:1991xy}
X.-N. Wang and M.~Gyulassy,
\newblock Phys.Rev.Lett. {\bf 68}, 1480 (1992).

\bibitem{Gyulassy:2003mc}
M.~Gyulassy, I.~Vitev, X.-N. Wang, and B.-W. Zhang,
\newblock (2003), arXiv:nucl-th/0302077.

\bibitem{Burke:2013yra}
JET, K.~M. Burke {\em et~al.},
\newblock Phys.Rev. {\bf C90}, 014909 (2014), arXiv:1312.5003.

\bibitem{Kang:2014xsa}
Z.-B. Kang, R.~Lashof-Regas, G.~Ovanesyan, P.~Saad, and I.~Vitev,
\newblock Phys.Rev.Lett. {\bf 114}, 092002 (2015), arXiv:1405.2612.

\bibitem{He:2011pd}
Y.~He, I.~Vitev, and B.-W. Zhang,
\newblock Phys.Lett. {\bf B713}, 224 (2012), arXiv:1105.2566.

\bibitem{Aad:2012vca}
ATLAS, G.~Aad {\em et~al.},
\newblock Phys.Lett. {\bf B719}, 220 (2013), arXiv:1208.1967.

\bibitem{Dokshitzer:2001zm}
Y.~L. Dokshitzer and D.~Kharzeev,
\newblock Phys.Lett. {\bf B519}, 199 (2001), arXiv:hep-ph/0106202.

\bibitem{Adare:2006nq}
PHENIX, A.~Adare {\em et~al.},
\newblock Phys.Rev.Lett. {\bf 98}, 172301 (2007), arXiv:nucl-ex/0611018.

\bibitem{Abelev:2006db}
STAR, B.~Abelev {\em et~al.},
\newblock Phys.Rev.Lett. {\bf 98}, 192301 (2007), arXiv:nucl-ex/0607012.

\bibitem{Wicks:2005gt}
S.~Wicks, W.~Horowitz, M.~Djordjevic, and M.~Gyulassy,
\newblock Nucl.Phys. {\bf A784}, 426 (2007), arXiv:nucl-th/0512076.

\bibitem{Adil:2006ra}
A.~Adil and I.~Vitev,
\newblock Phys.Lett. {\bf B649}, 139 (2007), arXiv:hep-ph/0611109.

\bibitem{vanHees:2007me}
H.~van Hees, M.~Mannarelli, V.~Greco, and R.~Rapp,
\newblock Phys.Rev.Lett. {\bf 100}, 192301 (2008), arXiv:0709.2884.

\bibitem{Qin:2009gw}
G.-Y. Qin and A.~Majumder,
\newblock Phys.Rev.Lett. {\bf 105}, 262301 (2010), arXiv:0910.3016.

\bibitem{Gossiaux:2008jv}
P.~Gossiaux and J.~Aichelin,
\newblock Phys.Rev. {\bf C78}, 014904 (2008), arXiv:0802.2525.

\bibitem{Sharma:2009hn}
R.~Sharma, I.~Vitev, and B.-W. Zhang,
\newblock Phys.Rev. {\bf C80}, 054902 (2009), arXiv:0904.0032.

\bibitem{ALICE:2012ab}
ALICE, B.~Abelev {\em et~al.},
\newblock JHEP {\bf 1209}, 112 (2012), arXiv:1203.2160.

\bibitem{Adare:2015kwa}
A.~Adare {\em et~al.},
\newblock (2015), arXiv:1501.06197.

\bibitem{Huang:2013vaa}
J.~Huang, Z.-B. Kang, and I.~Vitev,
\newblock Phys.Lett. {\bf B726}, 251 (2013), arXiv:1306.0909.

\bibitem{Chatrchyan:2013exa}
CMS, S.~Chatrchyan {\em et~al.},
\newblock Phys.Rev.Lett. {\bf 113}, 132301 (2014), arXiv:1312.4198.

\bibitem{Stavreva:2009vi}
T.~Stavreva and J.~Owens,
\newblock Phys.Rev. {\bf D79}, 054017 (2009), arXiv:0901.3791.

\bibitem{Hartanto:2013aha}
H.~Hartanto and L.~Reina,
\newblock Phys.Rev. {\bf D89}, 074001 (2014), arXiv:1312.2384.

\bibitem{Kang:2011rt}
Z.-B. Kang and I.~Vitev,
\newblock Phys.Rev. {\bf D84}, 014034 (2011), arXiv:1106.1493.

\bibitem{Stavreva:2012aa}
T.~Stavreva, F.~Arleo, and I.~Schienbein,
\newblock JHEP {\bf 1302}, 072 (2013), arXiv:1211.6744.

\bibitem{Vitev:2006bi}
I.~Vitev, J.~T. Goldman, M.~Johnson, and J.~Qiu,
\newblock Phys.Rev. {\bf D74}, 054010 (2006), arXiv:hep-ph/0605200.

\bibitem{Zhu:2006er}
X.~Zhu {\em et~al.},
\newblock Phys.Lett. {\bf B647}, 366 (2007), arXiv:hep-ph/0604178.

\bibitem{Cao:2015cba}
S.~Cao, G.-Y. Qin, and S.~A. Bass,
\newblock (2015), arXiv:1505.01869.

\bibitem{Sjostrand:2007gs}
T.~Sjostrand, S.~Mrenna, and P.~Z. Skands,
\newblock Comput.Phys.Commun. {\bf 178}, 852 (2008), arXiv:0710.3820.

\bibitem{Pumplin:2002vw}
J.~Pumplin {\em et~al.},
\newblock JHEP {\bf 0207}, 012 (2002), arXiv:hep-ph/0201195.

\bibitem{Cacciari:2008gp}
M.~Cacciari, G.~P. Salam, and G.~Soyez,
\newblock JHEP {\bf 0804}, 063 (2008), arXiv:0802.1189.

\bibitem{Abazov:2012ea}
D0, V.~Abazov {\em et~al.},
\newblock Phys.Lett. {\bf B714}, 32 (2012), arXiv:1203.5865.

\bibitem{Aaltonen:2013coa}
CDF, T.~Aaltonen {\em et~al.},
\newblock Phys.Rev.Lett. {\bf 111}, 042003 (2013), arXiv:1303.6136.

\bibitem{ATLAS:2011ac}
ATLAS, G.~Aad {\em et~al.},
\newblock Eur.Phys.J. {\bf C71}, 1846 (2011), arXiv:1109.6833.

\bibitem{Markert:2008jc}
C.~Markert, R.~Bellwied, and I.~Vitev,
\newblock Phys.Lett. {\bf B669}, 92 (2008), arXiv:0807.1509.

\bibitem{Neufeld:2011yh}
R.~Neufeld and I.~Vitev,
\newblock Phys.Rev. {\bf C86}, 024905 (2012), arXiv:1105.2067.

\bibitem{Neufeld:2014yaa}
R.~Neufeld, I.~Vitev, and H.~Xing,
\newblock Phys.Rev. {\bf D89}, 096003 (2014), arXiv:1401.5101.

\bibitem{Neufeld:2012df}
R.~Neufeld and I.~Vitev,
\newblock Phys.Rev.Lett. {\bf 108}, 242001 (2012), arXiv:1202.5556.

\bibitem{Dai:2012am}
W.~Dai, I.~Vitev, and B.-W. Zhang,
\newblock Phys.Rev.Lett. {\bf 110}, 142001 (2013), arXiv:1207.5177.

\bibitem{d'Enterria:2003qs}
D.~G. d'Enterria,
\newblock (2003), arXiv:nucl-ex/0302016,
\newblock see also: \url {http://dde.web.cern.ch/dde/glauber_lhc.htm}.

\bibitem{Vitev:2008rz}
I.~Vitev, S.~Wicks, and B.-W. Zhang,
\newblock JHEP {\bf 0811}, 093 (2008), arXiv:0810.2807.

\bibitem{Chatrchyan:2012gt}
CMS, S.~Chatrchyan {\em et~al.},
\newblock Phys.Lett. {\bf B718}, 773 (2013), arXiv:1205.0206.

\bibitem{Chatrchyan:2011sx}
CMS, S.~Chatrchyan {\em et~al.},
\newblock Phys.Rev. {\bf C84}, 024906 (2011), arXiv:1102.1957.

\bibitem{Aad:2010bu}
ATLAS, G.~Aad {\em et~al.},
\newblock Phys.Rev.Lett. {\bf 105}, 252303 (2010), arXiv:1011.6182.

\bibitem{Ovanesyan:2011kn}
G.~Ovanesyan and I.~Vitev,
\newblock Phys.Lett. {\bf B706}, 371 (2012), arXiv:1109.5619.

\bibitem{Kang:2013raa}
Z.-B. Kang, E.~Wang, X.-N. Wang, and H.~Xing,
\newblock Phys.Rev.Lett. {\bf 112}, 102001 (2014), arXiv:1310.6759.

\bibitem{Kang:2014ela}
Z.-B. Kang, E.~Wang, X.-N. Wang, and H.~Xing,
\newblock (2014), arXiv:1409.1315.

\end{thebibliography}

\end{document}